\documentclass[11pt,a4paper]{article}
\usepackage{jheppub}

\usepackage{hyperref}  

\newcommand{\vev}[1]{\langle #1 \rangle}

\begin{document}
\title{Spontaneous parity breaking and supersymmetry breaking in metastable vacua 
with consistent cosmology  
\\}
\author{Debasish Borah}
\emailAdd{debasish@phy.iitb.ac.in}
\author{and Urjit A. Yajnik}
\emailAdd{yajnik@phy.iitb.ac.in}
\affiliation{Department of Physics, Indian Institute of Technology Bombay,  \\
Mumbai 400076, India}

\abstract{
We study the compatibility of spontaneous breaking of parity and  successful cosmology
in a left-right symmetric model where supersymmetry breaking is achieved in metastable vacua.
We show that domain walls formed due to this breaking can be removed due to 
Planck scale  suppressed terms,  provided the parity breaking scale $M_R$ 
is constrained to remain smaller than $10^{10}-10^{11}$ GeV. Ensuring metastability is achieved
naturally even if the entire mechanism operates at low scales, within a few orders of magnitude
of the TeV scale. Taking $M_R$ as high as permitted, close to the acceptable reheat temperature
after inflation,  would require the  magnetic phase of the  Supersymmetric Quantum Chromodynamics (SQCD) to have set in before the 
end of inflation.
}
\maketitle

\section{Introduction}

Left-Right Symmetric Models (LRSM) \cite{Pati:1974yy,Mohapatra:1974gc, Senjanovic:1975rk, 
Mohapatra:1980qe, Deshpande:1990ip} with the gauge symmetry $SU(3)_c \times SU(2)_L \times SU(2)_R \times U(1)_{B-L}$ 
provide a framework within which spontaneous parity breaking as well as tiny neutrino 
masses \cite{Fukuda:2001nk, Ahmad:2002jz, 
Ahmad:2002ka, Bahcall:2004mz} can be successfully implemented without reference to very high 
scale physics such as grand unification. The structure of the gauge group also suggests a discrete 
$Z_2$ symmetry, so called D-parity \cite{Chang:1983fu,Chang:1984uy} under which the left and right 
handed fields get interchanged and the gauge charges $g_{2L}$, $g_{2R}$ are equal at some suitable scale. 
In supersymmetric implementations of Left-Right symmetry 
\cite{Aulakh:1998nn, Aulakh:1997ba,Babu:2008ep,Patra:2009wc}, stability of gauge
hierarchy becomes natural and R-parity $R_p =(-1)^{3(B-L)+2S}$ (where $S$ is the spin,) 
is automatically a part of the  gauge symmetry. A stable Dark Matter candidate is therefore more
natural than in the Minimal Supersymmetric Standard Model.

Spontaneous breaking of exact discrete symmetries  has cosmological
implications since they lead to frustrated phase transitions leaving behind a 
network of domain walls. These domain walls, if not removed will be in conflict 
with the observed Universe \cite{Kibble:1980mv,Hindmarsh:1994re}. 
It was pointed out \cite{Rai:1992xw,Lew:1993yt} that Planck scale suppressed non-renormalizable 
operators can be a source of domain wall instability. Interestingly, this generic
analysis needs a careful revision when supersymmetry (SUSY) and gauge symmetries need
to be incorporated. In the Next to Minimal Supersymmetric Standard Model, the problem 
was found to persist \cite{Abel:1995wk}, in the sense that the gauge hierarchy problem 
does not get addressed if the operators required to remove the domain walls
are permitted. The problems encountered in that model are however generic to
introduction of a gauge singlet. In the Supersymmetric Left-Right Models (SUSYLR)
with all Higgs carrying gauge charges, it is possible to introduce Planck scale suppressed 
terms that are well regulated. One can then demand that the new operators ensure 
sufficient pressure across the domain walls that the latter disappear
before Big Bang Nucleosynthesis (BBN). 
This requirement has been discussed in \cite{Mishra:2009mk} in the context of $R$-parity  
conserving SUSYLR models ~\cite{Aulakh:1998nn, Aulakh:1997ba,Babu:2008ep}. 
Similar analysis was shown to place constraints 
also on R-parity violating SUSYLR models \cite{Borah:2011qq}. The upshot is that the
framework proposed in  \cite{Mishra:2009mk} gives rise to an upper bound on the D-parity 
breaking scale. And this scale is found to be $\sim 10^{10}$GeV, far below the grand unification 
scale. 
The issue of how supersymmetry breaking may be achieved in these models is
open and deserves special attention. In \cite{Mishra:2008be} gauge mediation mechanism
was tailored to left-right symmetric case. It was however found to be unnatural to
expect the $D$-parity breaking operators to also emerge from the hidden sector.

Here we are interested in models where SUSY breaking is achieved without any strongly coupled 
hidden sector. Such models are based on the idea that SUSY breaking vacuum is a 
long lived metastable vacuum, as originally proposed by Intriligator, Seiberg and Shih (ISS) \cite{Intriligator:2006dd}. 
This approach to SUSY breaking for the Minimal Supersymmetric Standard Model 
was pursued in \cite{Abel:2007uq} and for $SO(10)$ in \cite{Koschade:2009qu}. 
Recently ISS type SUSY breaking was considered for Left-Right symmetric model by Haba and Ohki \cite{Haba:2011pr}. 
However, $D$-parity is broken only locally in this model, so that in the early Universe
there would be  formation of patches corresponding to different vacua separated by a network 
of domain walls. We take the model \cite{Haba:2011pr} as a specific realization and study the domain 
wall disappearance using the framework proposed in \cite{Mishra:2009mk}.

The outcome of this study is that as in previous studies, an upper bound is required
on the scale of the $D$-parity breaking and therefore the scale $M_R$ of $SU(2)_R$
breaking. As a simple possibility, the entire program is successfully implemented if 
we treat all of the new physics to be within a few orders of magnitude of the known 
scale $\sim$TeV, but in this case gauge coupling unification would be problematic \cite{Borah:2010kk}.
An alternative is to note that the bound is tantalizingly close to the intermediate scale 
advocated in some string theoretic unification models ( see for instance, \cite{LEstrings1} \cite{LEstrings2}).
But this scale is numerically comparable to the bound on the reheat temperature
$T_{RH}$ after inflation required for the avoidance of gravitino overproduction. In the context of 
implementing direct supersymmetry breaking, this has implications also for the scale
$\Lambda_m$ below which the underlying SQCD can be treated as a magnetic theory, and its
relation to inflation, as discussed in the concluding section.
 
Before proceeding, it is worth mentioning that there are models where $D$-parity and 
$SU(2)_R$ gauge symmetry  are broken at two different stages, at the cost of introduction 
of a gauge singlet scalar field. Such models do not suffer from the problem of 
persistent domain walls \cite{Chang:1983fu,Chang:1984uy,Sarkar:2007er,Patra:2009wc,Dev:2009aw,Borah:2010zq,Borah:2011zz}.
If the model is descended from breaking of $SO(10)$, there is an
interesting alternative to be studied. Firstly we observe that the
breaking pattern of $SO(10)$ is model dependent and exact left-right
symmetry may not be an effective symmetry at any of the lower energy
scales. However, in models where an exact left-right symmetry occurs
naturally in an intermediate energy regime, domain walls are bound
to occur when the $D$ parity as an effective discrete symmetry gets
broken. However, in this case Planck suppressed  terms ensuring the
disappearance of the domain walls do not occur, because being a
gauge symmetry, quantum gravity effects do not naturally break it.
This case however  is not pursued here.

This paper is organized as follows. In section \ref{sec:DW-section} we briefly review the 
domain wall dynamics. In section \ref{sec:ISS1} we summarize the ISS model and in section \ref{sec:ISSParity} 
we discuss the model originally proposed by Haba and Ohki \cite{Haba:2011pr} and then discuss how gravity 
can cure the problem of domain walls in this model in section \ref{sec:dwallremove}. We also discuss the longevity of the metastable vacua in section \ref{sec:longevity} and summarize our results in 
section \ref{sec:con}.

\section{Domain Wall Dynamics}
\label{sec:DW-section}
Discrete symmetries and their spontaneous breaking are both common instances and desirable in 
model building. The spontaneous breaking of such discrete symmetries gives rise to a network of 
domain walls leaving the accompanying phase transition frustrated \cite{Kibble:1980mv, Hindmarsh:1994re}. 
The danger of a frustrated phase transition can therefore be evaded if a small explicit breaking of discrete 
symmetry can be introduced.

Due to the smallness of such discrete symmetry breaking, the resulting domain walls may be relatively long lived and
 can dominate the Universe for a long time. Since this will be in conflict with the observed Universe, these domain 
walls need to disappear at a very high energy scale (at least before BBN). Keeping this in mind, we summarize the two cases 
of domain wall dynamics discussed in \cite{Mishra:2009mk}, one in which the domain walls originate in radiation dominated era and get destabilized also within
the radiation dominated era. This scenario was originally considered by Kibble \cite{Kibble:1980mv} and Vilenkin \cite{Vilenkin:1984ib}. The second scenario was
essentially proposed in \cite{Kawasaki:2004rx}, which consists of the walls
originating  in a radiation dominated phase, but decaying after the 
Universe enters a matter dominated phase, either due to substantial 
production of heavy unwanted relics such as moduli, or simply due
to a coherent oscillating scalar field. In both the cases the domain
walls disappear before they come to dominate the energy density of the Universe.

When a scalar field $\phi$ acquires a vev at a scale $M_R$ at some critical temperature $T_c$, a phase transition occurs leading to the formation of domain walls. The energy density trapped per unit area of such a wall is $\sigma \sim  M^3_R$. The dynamics of the walls are determined by two quantities, force due to tension $f_T \sim \sigma/R$ and force due to friction $f_F \sim \beta T^4$ where $R$ is the average scale of radius of curvature prevailing in the wall complex, $\beta$ is the speed at which the domain wall is navigating through the medium and $T$ is the temperature. The epoch at which these two forces balance each other sets the time scale $t_R \sim R/\beta $. Putting all these together leads to the scaling law for the growth of the scale $R(t)$:
\begin{equation}
R(t) \approx (G \sigma)^{1/2}t^{3/2}
\end{equation}
The energy density of the domain walls goes as $\rho_W \sim (\sigma R^2/R^3) \sim (\sigma/Gt^3)^{1/2} $. In a radiation dominated era this $\rho_W$ is comparable to the energy density of the Universe $[\rho \sim 1/(Gt^2)]$ around time $t_0 \sim 1/(G \sigma)$. 

The pressure difference arising from small asymmetry on the two sides of the wall competes with the two forces $f_F \sim 1/(Gt^2)$ and $f_T \sim (\sigma/(Gt^3))^{1/2}$ discussed above. For $\delta \rho$ to exceed either of these two quantities before $t_0 \sim 1/(G\sigma)$
\begin{equation}
 \delta \rho \geq G\sigma^2 \approx \frac{M^6_R}{M^2_{Pl}} \sim M_R^4 \left(\frac{M_R}{M_{Pl}}\right)^2
\label{eq:RD-delta-rho}
\end{equation}

Similar analysis in the matter dominated era, originally considered in \cite{Kawasaki:2004rx} begins with the assumption that the initially formed wall complex in a phase transition is expected to rapidly relax to a few walls per horizon volume at an epoch characterized by Hubble parameter value $H_i$. Thus the initial energy density of the wall complex is $\rho^{\text{in}}_W \sim \sigma H_i$. This epoch onward the energy density of the Universe is assumed to be dominated by heavy relics or an oscillating modulus field and in both the cases the scale factor grows as $a(t) \propto t^{2/3}$. The energy density scales as $\rho_{\text{mod}} \sim \rho^{\text{in}}_{\text{mod}}/(a(t))^3$. If the domain wall (DW) complex remains frustrated, i.e. its energy density contribution $\rho_{\text{DW}} \propto 1/a(t)$, the Hubble parameter at the epoch of equality of DW contribution with that of the rest of the matter is given by \cite{Kawasaki:2004rx}
\begin{equation}
H_{\text{eq}} \sim \sigma^{3/4}H^{1/4}_i M^{-3/2}_{Pl}
\label{heq}
\end{equation}
Assuming that the domain walls start decaying as soon as they dominate the energy density of the Universe, which corresponds to a temperature $T_D$ such that $H^2_{\text{eq}} \sim GT^4_D$, the above equation gives 
\begin{equation}
T^4_D \sim \sigma^{3/2} H^{1/2}_i M^{-1}_{Pl}
\label{td}
\end{equation}
Under the assumption that the domain walls are formed at $T \sim \sigma^{1/3}$
\begin{equation}
H^2_i = \frac{8\pi}{3}G\sigma^{4/3} \sim \frac{\sigma^{4/3}}{M^2_{Pl}}
\end{equation}
Now from Eq. (\ref{td})
\begin{equation}
T^4_D \sim \frac{\sigma^{11/6}}{M^{3/2}_{Pl}} \sim \frac{M^{11/2}_R}{M^{3/2}_{Pl}} \sim M^4_R \left(\frac{M_R}{M_{Pl}}\right)^{3/2}
\end{equation}
Demanding $\delta \rho > T^4_D$ leads to
\begin{equation}
\delta \rho >  M^4_R \left(\frac{M_R}{M_{Pl}}\right)^{3/2}
\end{equation}

\section{ISS Model: A Recap}
\label{sec:ISS1}
The ISS model consists of a pair of dual theories related to each 
other through the ``Seiberg duality'' \cite{Seiberg:1994pq}. 
There is a low energy theory which is referred to as the ``macroscopic''
or ``free magnetic theory'' which is infra-red (IR) free. The high energy theory
is known as the ``microscopic'' or ``free electric theory'' and it 
is basically $SU(N_c)$ SQCD which is ultra violet (UV) free. Seiberg duality says, $SU(N_c)$ SQCD (UV free) with $N_f (>N_c)$ flavors of quarks is dual to a $SU(N_f-N_c)$ gauge theory (IR free) with $N^2_f$ singlet mesons $M$ and $N_f$ flavors of quarks $q, \tilde{q}$. Their approach revolves around studying the SUSY-breaking
dynamics of the microscopic $SU(N_c)$ SQCD in terms of the
macroscopic, IR-free dual. 

\subsection{The macroscopic model}

The macroscopic model is a Wess-Zumino model with the following symmetry group out of which $SU(N)$ where $N= N_f-N_c$ is gauged and the rest are global symmetries:
\begin{equation}
SU(N) \times SU(N_f)^2 \times U(1)_B \times U(1)' \times U(1)_R.
\end{equation}
The matter content is as follows:
\begin{equation}
 \begin{matrix}
        & SU(N) & SU(N_f) & SU(N_f) & U(1)_B & U(1)' & U(1)_R \cr
        & \cr
       \Phi_{[N_f \times N_f]} & 1 & \square  & \overline{\square} &  0 & -2 &  2\cr
       & \cr
       \varphi_{[N \times N_f]} & \square & \overline{\square} & 1 & 1 & 1 & 0 \cr
       & \cr
       \tilde \varphi_{[N_f \times N]} & \overline{\square} & 1 & \square & -1 & 1 & 0\cr         
 \end{matrix}
\end{equation}
The field $\Phi$ is identified as the meson field. The fields $\varphi$ and $\tilde{\varphi}$ are 
the dual quarks. The K$\ddot{a}$hler potential is
\begin{equation}
K = {\rm Tr}\left[\varphi^{\dagger}\varphi\right] +
 {\rm Tr}\left[\tilde{\varphi}^{\dagger}\tilde{\varphi}\right] +
 {\rm Tr}\left[\Phi^{\dagger}\Phi\right]
\end{equation}
and the tree-level superpotential is
\begin{equation}
W = h{\rm Tr}\left[\varphi\Phi\tilde{\varphi}\right] -
 h\mu^2{\rm Tr}\Phi.
\end{equation}
Denoting the scalar components of the superfields $\Phi, \phi, \tilde{\phi}$ as $M$, $q$, and $\tilde{q}$ respectively, the bosonic part of the 
Lagrangian for macroscopic theory can be expressed as 
\begin{equation}
{\cal L} = {\rm Tr}_c
\left[ - \frac{1}{2g^2} F_{\mu\nu} F^{\mu\nu}
- {\cal D}_\mu q {\cal D}^\mu q^\dagger
- {\cal D}_\mu \widetilde q^\dagger {\cal D}^\mu \widetilde q
\right]
- {\rm Tr}_f \left[
\partial_\mu M^\dagger \partial^\mu M
\right]
- V
\end{equation}
with a scalar potential $V = V_F + V_D$ given by
\begin{eqnarray}
V_F &=& |h^2|
{\rm Tr}_f \left[
\left|\widetilde q q - \mu^2 {\bf 1}_{N_f} \right|^2
\right] +\ |h^2| {\rm Tr}_c \left[
\left| q M \right|^2
+ \left| \widetilde q^\dagger M^\dagger \right|^2
\right],
\label{iss_pot_1}\\
V_D &=& \frac{g^2}{4} {\rm Tr}_c
\left[
\left(q q^\dagger - \widetilde q^\dagger \widetilde q\right)^2
\right]
-\frac{g^2}{8} 
\left({\rm Tr}_c q q^\dagger - 
{\rm Tr}_c\widetilde q^\dagger \widetilde q\right)^2.
\label{iss_pot_2}
\end{eqnarray}

\subsection{SUSY breaking vacua}

SUSY is broken in the above model by rank condition if
$N < N_f$. The most general SUSY breaking vacuum is of the form 
\begin{equation}
M =
\left(
\begin{array}{cc}
0_{N \times N} & 0_{N \times (N_f-N)}\\
0_{(N_f-N) \times N} & M_{0}
\end{array}
\right),\quad
\frac{q^\dagger}{\mu^*} = \frac{\widetilde q }{\mu}
=  \left(
\begin{array}{c}
{\bf 1}_N\\
0_{(N_f-N) \times N}
\end{array}
\right),
\label{VEV}
\end{equation}
where $M_0$ is an arbitrary $N_f - N$ by $N_f - N$ matrix. The vacuum energy is given by
\begin{equation}
V = |h \mu^2|^2 (N_f - N) > 0,
\label{meta}
\end{equation}
and the vacuum spontaneously breaks the ${\cal N}=1$ supersymmetry.

\subsection{SUSY vacuum}

From equations (\ref{iss_pot_1}) and (\ref{iss_pot_2}), the supersymmetric vacua lie at
\begin{equation}
\widetilde q q = \mu^2 {\bf 1}_{N_f},\quad
q M = 0,\quad
\widetilde q^\dagger M^\dagger = 0,\quad
q q^\dagger - \widetilde q^\dagger \widetilde q = 0.
\label{vacuum}
\end{equation}
For these vacua, $q$ and $\tilde{q}$ are non-zero and hence 
$M$ must be zero. However if we give
$M$ some general non-zero VEV $\vev{M}$ as in equation (\ref{VEV}), the dual 
quarks $\varphi$ and $\tilde{\varphi}$ acquire a mass $h\vev{M}$.
If we integrate out these massive flavors of quarks, the low energy superpotential
becomes 
\begin{equation}
W_{\text{low}} = N(h^{N_f}\Lambda^{-(N_f-3N)}_m \text{det} \Phi )^{1/N} -h\mu^2 \text{Tr} \Phi
\end{equation}
Minimizing the above superpotential gives rise to the supersymmetric minima at
\begin{equation}
 \langle h M\rangle = \Lambda _m\epsilon ^{2N/(N_f-N)}{\bf 1}
 _{N_f}=\mu {\frac{1}{\epsilon^{(N_f-3N)/(N_f-N)}}}{\bf 1}_{N_f}
 \label{susyminima}
\end{equation}
where $\epsilon \equiv {\frac{\mu}{\Lambda _m}}$.
The value of $\varphi$ and $\tilde{\varphi}$ or equivalently $q$ 
and $\tilde{q}$ for these minima is zero. $\Lambda_m$ refers to the
scale above which the magnetic theory is strongly coupled. For $\lvert \epsilon \rvert \ll 1 $, the SUSY preserving vacuum lies far away in the field space from the SUSY breaking vacuum and hence can be parametrically long lived as was pointed out by Intriligator et al. \cite{Intriligator:2006dd}.

\section{D-Parity breaking with ISS type SUSY breaking}
\label{sec:ISSParity}
As suggested by the authors in \cite{Haba:2011pr}, D-parity and spontaneous SUSY breaking can be naturally achieved in an ISS type framework if a new strongly coupled gauge sector is introduced to which the left-right Higgs fields are coupled. They proposed the electric gauge theory to be based on the gauge group $SU(3)_L \times SU(3)_R \times SU(2)_L \times SU(2)_R \times U(1)_{B-L}$ (in short $G_{33221}$) where $SU(2)_L \times SU(2)_R \times U(1)_{B-L}$ is the gauge group of usual Left-Right models and $SU(3)_{L,R}$ is the new strongly coupled gauge sector introduced. The dual description similar to the original ISS model gives rise to $SU(2)_R$ broken meta-stable vacua inducing spontaneous SUSY breaking simultaneously. 

The particle content of the electric theory is
$$ Q_L^a \sim (3,1,2, 1, 1), \quad \tilde{Q}^a_L \sim (3^*, 1, 2, 1, -1) $$
$$ Q_R^a \sim (1,3,1, 2, -1), \quad \tilde{Q}^a_R \sim (1,3^*,1, 2, 1) $$
where $a = 1, N_f$ and the numbers in brackets correspond to the transformations of the fields under the gauge group $G_{33221}$. This model has $N_c = 3$ and hence to have a Seiberg dual \cite{Seiberg:1994pq} magnetic theory, number of flavors should be $N_f \geq 4$. For $N_f = 4$ the dual magnetic theory will have the gauge symmetry of the usual Left Right Models $SU(2)_L \times SU(2)_R \times U(1)_{B-L}$ and the following particle content
$$\phi^a_L (2,1,-1), \quad \tilde{\phi}^a_L (2,1,1) $$
 $$\phi^a_R (1,2,1), \quad \tilde{\phi}^a_R (1,2,-1) $$
\begin{equation}
\Phi_L \equiv \bf{1}+\text{Adj}_L =
\left(\begin{array}{cc}
\ \frac{1}{\surd 2}(S_L+\delta^0_L) & \delta^+_L \\
\ \delta^-_L & \frac{1}{\surd 2}(S_L-\delta^0_L )
\end{array}\right) \nonumber
\end{equation}
\begin{equation}
\Phi_R \equiv \bf{1}+\text{Adj}_R =
\left(\begin{array}{cc}
\ \frac{1}{\surd 2}(S_R+\delta^0_R) & \delta^+_R \\
\ \delta^-_R & \frac{1}{\surd 2}(S_R-\delta^0_R )
\end{array}\right)
\label{eq:deltaCcomponents}
\end{equation}
The Left-Right symmetric renormalizable superpotential of this magnetic theory is 
\begin{equation}
W^0_{LR} = h \text{Tr} \phi_L \Phi_L \tilde{\phi}_L -h \mu^2 \text{Tr} \Phi_L+h \text{Tr} \phi_R \Phi_R \tilde{\phi}_R -h \mu^2 \text{Tr} \Phi_R
\label{WLR1}
\end{equation}
The tree level K$\ddot{a}$hler potential is
\begin{equation}
K_0 = \text{Tr}\phi^{\dagger}_L \phi_L + \text{Tr} \tilde{\phi}^{\dagger}_L \tilde{\phi}_L+\text{Tr}\phi^{\dagger}_R \phi_R + \text{Tr} \tilde{\phi}^{\dagger}_R \tilde{\phi}_R + \text{Tr}\Phi^{\dagger}_L \Phi_L+\text{Tr}\Phi^{\dagger}_R \Phi_R
\end{equation}
The non-zero F-terms giving rise to SUSY breaking are
\begin{equation}
F_{\Phi_L} = h \phi_L \tilde{\phi}_L -h \mu^2 \delta_{ab} \qquad \mathrm{and}
 F_{\Phi_R} = h \phi_R \tilde{\phi}_R -h \mu^2 \delta_{ab} 
\end{equation}
where $a,b = 1, 4$ here and SUSY is broken by rank condition \cite{Intriligator:2006dd}. However depending on the vev of the meson fields $\langle \Phi_L \rangle, \langle \Phi_R \rangle$, the chiral fields $\phi_L, \phi_R$ will acquire different masses proportional to $h \langle \Phi_L \rangle $ and $h \langle \Phi_R \rangle $ respectively. Suppose, $\Phi_R$ gets a non-zero vev and accordingly $\phi_R$ acquire non-zero masses. Like in the ISS model (\ref{susyminima}) is we integrate out these massive flavors, we arrive at the SUSY preserving vacuum with only the left handed chiral fields $\phi_L$. 
After integrating out the right handed chiral fields, the superpotential becomes
\begin{equation}
W^0_{L} = h \text{Tr} \phi_L \Phi_L \tilde{\phi}_L -h \mu^2 \text{Tr} \Phi_L+ h^4 \Lambda^{-1} \text{det}\Phi_R -h \mu^2 \text{Tr} \Phi_R
\label{WL1}
\end{equation}
which gives rise to SUSY preserving vacua at 
\begin{equation}
\langle h \Phi_R \rangle = \Lambda_m \epsilon^{2/3} = \mu \frac{1}{\epsilon^{1/3}}
\label{rightvev}
\end{equation}
where $\epsilon = \frac{\mu}{\Lambda_m}$.
Thus the right handed sector exists in a metastable SUSY breaking vacuum whereas the left handed sector is in a SUSY preserving vacuum breaking D-parity spontaneously. Soft SUSY breaking terms can also be induced in the right handed sector as pointed out by the authors of \cite{Haba:2011pr}. However it is equally likely for $\Phi_L$ to acquire a vev instead of $\Phi_R$. In that case left handed fields will acquire a mass $h \langle \Phi_L \rangle $ and get decoupled. Integrating them out will give rise to a SUSY preserving right handed sector. Thus D-parity is broken only locally and there will be formation of local patches containing left and right handed sectors separated by a network of domain walls. Domain walls, being extended objects have more energy density than matter and radiation and hence start dominating the Universe very early. This will be in conflict with the observed Universe and hence such walls should be removed at early times or at least before BBN. We discuss one such possible way to get rid of them in the next section.

\section{Domain Wall Removal}
\label{sec:dwallremove}
Although the electric theory is Left-Right symmetric, the magnetic theory breaks both D-parity as well as SUSY spontaneously as discussed in the previous section. D-parity is broken only locally giving rise to the formation of domain walls. Following \cite{Rai:1992xw,Lew:1993yt}, we know that Planck scale suppressed operators can break D-parity explicitly and can make the domain walls disappear. However the magnetic theory where D-parity is broken, has an UV cut-off $\Lambda_m$ and hence all the gauge invariant higher dimensional terms will be suppressed by $\Lambda_m$ and not by Planck mass $M_{Pl}$. But QCD effects can not be responsible for breaking D-parity although quantum gravity effects can break global discrete symmetries like D-parity explicitly. Therefore we assume that the differences in the left and right sectors brought about by $\Lambda_m$ suppressed operators are of the order $\frac{1}{M_{Pl}}$.

We write the next to leading order terms allowed by the gauge symmetry in the superpotential as well as K$\ddot{a}$hler potential.
\begin{equation}
W^1_{LR} = f_L \frac{\text{Tr}(\phi_L \Phi_L \tilde{\phi}_L) \text{Tr} \Phi_L}{\Lambda_m} +f_R \frac{\text{Tr}(\phi_R \Phi_R \tilde{\phi}_R) \text{Tr} \Phi_R}{\Lambda_m}+ f'_L \frac{(\text{Tr} \Phi_L )^4}{\Lambda_m}+f'_R \frac{(\text{Tr} \Phi_R )^4}{\Lambda_m}
\end{equation}
$$ K_1 = -\lambda_{1L} \frac{\text{Tr}(\Phi^{\dagger}_L \Phi_L)^2}{\Lambda^2_m}-\lambda_{2L} \frac{(\text{Tr}\Phi^{\dagger}_L\Phi_L )^2}{\Lambda^2_m}-\lambda_{1R} \frac{\text{Tr}(\Phi^{\dagger}_R \Phi_R)^2}{\Lambda^2_m}-\lambda_{2R} \frac{(\text{Tr}\Phi^{\dagger}_R\Phi_R )^2}{\Lambda^2_m} $$
\begin{equation}
 -\frac{\lambda'_{1L}}{\Lambda^2_m} ((\phi^{\dagger}_L\phi_L)^2+(\tilde{\phi}^{\dagger}_L \tilde{\phi}_L)^2)-\frac{\lambda'_{1R}}{\Lambda^2_m} ((\phi^{\dagger}_R\phi_R)^2+(\tilde{\phi}^{\dagger}_R \tilde{\phi}_R)^2)
\end{equation}
After the right(left) sector decouples we are left with the left(right) sector only. We find the energy of these two sectors separately. The terms of order $\frac{1}{\Lambda_m}$ are given by
\begin{equation}
V^1_R =  \frac{h}{\Lambda_m} S_R [f_R(\phi^0_R \tilde{\phi}^0_R)^2+f'_R\phi^0_R \tilde{\phi}^0_R S^2_R +(\delta^0_R -S_R)^2((\phi^0_R)^2+(\tilde{\phi}^0_R)^2)]
\end{equation}
The minimization conditions give $\phi \tilde{\phi} = \mu^2 $ and $S^0 = -\delta^0$. Denoting $\langle \phi^0_R \rangle = \langle \tilde{\phi}^0_R \rangle = \mu $ and $\langle \delta^0_R \rangle =-\langle S^0_R \rangle = M_R$, we have 
\begin{equation}
V^1_R = \frac{hf_R}{\Lambda_m} (\lvert \mu \rvert^4 M_R +\lvert \mu \rvert^2 M^3_R ) 
\end{equation}
where we have also assumed $f'_R \approx f_R$. For $ \lvert \mu \rvert < M_R$ we have 
\begin{equation}
V^1_R = \frac{hf_R}{\Lambda_m} \lvert \mu \rvert^2 M^3_R
\end{equation}
If the scalar fields in the left sector also acquires similar vev, the terms of the order $\frac{1}{\Lambda_m}$ in the expression for energy are 
\begin{equation}
V^1_L =  \frac{hf_L}{\Lambda_m} \lvert \mu \rvert^2 M^3_R
\end{equation}
Thus the effective energy density is
\begin{equation}
\delta \rho \sim h(f_R-f_L) \frac{\lvert \mu \rvert^2 M^3_R}{\Lambda_m}
\end{equation}
Assuming quantum gravity effects to bring such an explicit violation of D-parity we must have 
\begin{equation}
\frac{\lvert \mu \rvert^2M^3_R}{\Lambda_m} \leq \frac{\Lambda^5_m}{M_{Pl}}
\label{MRMPL}
\end{equation}
Here we are considering the dimensionless coefficients to be of order one. The above relation implies $ \Lambda^6_m \geq M_{Pl}\lvert \mu \rvert^2M^3_R $. For matter dominated era, we must have 
\begin{equation}
h(f_R-f_L) \frac{\lvert \mu \rvert^2M^3_R}{\Lambda_m} > M^4_R \left(\frac{M_R}{M_{Pl}}\right)^{3/2}
\end{equation}
Assuming the dimensionless parameters to be of order one, the above relation gives 
\begin{equation}
M^{5/2}_R < \frac{\lvert \mu \rvert^2 M^{3/2}_{Pl}}{\Lambda_m}
\end{equation}
Using lower bound on $\Lambda_m$ from equation (\ref{MRMPL}), the above inequality gives the upper bound on $M_R$
\begin{equation}
M_R < \lvert \mu \rvert^{5/9} M^{4/9}_{Pl}
\end{equation}
Taking $\mu$ to be of same order as SUSY breaking scale which is TeV, we get
\begin{equation}
M_R < 1.3 \times 10^{10}\; \text{GeV}
\label{MRbound1}
\end{equation}
Similarly for radiation dominated era, we have
\begin{equation}
h(f_R-f_L) \frac{\lvert \mu \rvert^2M^3_R}{\Lambda_m} >\frac{M^6_R}{M^2_{Pl}} 
\end{equation}
Assuming dimensionless parameters to be of order one, we have
\begin{equation}
M^3_R < \frac{\lvert \mu \rvert^2 M^2_{Pl}}{\Lambda_m}
\end{equation}
Using lower bound on $\Lambda_m$ from equation (\ref{MRMPL}), the above inequality gives the upper bound on $M_R$
\begin{equation}
M_R < \lvert \mu \rvert^{10/21} M^{11/21}_{Pl}
\end{equation}
Taking $\mu$ to be of same order as SUSY breaking scale which is TeV, we get
\begin{equation}
M_R < 10^{11}\; \text{GeV}
\label{MRbound2}
\end{equation}
It is interesting to note that $M_R$ as constrained in (\ref{MRbound1}) is just above the upper bound on the reheat 
temperature $T_{RH}$ after inflation required to avoid gravitino overabundance. The constraint (\ref{MRbound2}) if 
saturated precludes the possibility of thermal leptogenesis \cite{Fukugita:1986hr} due to the gravitino bound on $T_{RH}$. 
But more importantly, the scale falls far short of either the scale of generic inflation $\sim 10^{15} - 10^{16} \; \text{GeV}$ \cite{Guo:2010mm,Boyle:2005ug,Komatsu:2008hk} or 
the scale of $SO(10)$ grand unification $\sim 2 \times 10^{16} \; \text{GeV}$. Thus the successful completion of D-parity 
breaking phase transition demands the introduction of this new scale lying below the grand unification scale. Within the 
framework of SUSY, such a scale can be assumed to be protected from mixing with higher scales.

\section{Longevity of metastable vacua}
\label{sec:longevity}
In the original ISS model \cite{Intriligator:2006dd}, the metastable non-supersymmetric vacua can be made parametrically 
long lived by taking the parameter $\epsilon \equiv \mu/\Lambda_m $ to be sufficiently small. In the semi-classical 
approximations \cite{Coleman:1977py} the tunneling probability from the metastable SUSY breaking vacua to the SUSY 
preserving true vacua is directly proportional to $\text{exp}(-S_b)$ where $S_b$ called the "bounce action" is the 
difference between the Euclidean action of the tunneling configuration and that of remaining in the metastable vacuum. 
The authors of \cite{Intriligator:2006dd} derived an approximate form of this bounce action using the techniques of 
\cite{Duncan:1992ai}
\begin{equation}
S_b = \frac{A}{\lvert \epsilon \rvert^{4(N_f-3N)/(N_f-N)}} 
\end{equation}
where $A$ is a factor of order one. The bounce action can be arbitrarily large and hence the metastable vacua can be 
arbitrarily long lived if $\epsilon \ll 1$. Similar analysis for the model of \cite{Haba:2011pr} will give rise 
to $S_b \sim 1/\lvert \epsilon \rvert^{4/3}$. For the metastable vacua to have a lifetime greater than the age of 
the Universe $(\sim 10^{18}\; \text{s})$ we have 
\begin{equation}
\text{exp}\{ \lvert \epsilon \rvert^{-4/3}\} > 10^{18}
\label{stabilitybound}
\end{equation}
which corresponds to the approximate bound $\epsilon  < 6 \times 10^{-2}$. Thus any value of $\epsilon$ 
below $10^{-2}$ 
will make the metastable vacua long lived. This bound is not significantly changed for modest
changes in the choice of $N_f$. From equation (\ref{rightvev}) 
we have $\langle h \Phi_R \rangle = \mu \frac{1}{\epsilon^{1/3}}$. Using $\mu \sim 1 \; \text{TeV}, h \approx 1$ and 
the above upper bound on $\epsilon$ we get the lower bound $M_R \sim \langle \Phi_R \rangle > 3 \;\text{TeV}$. 
Thus the Left-Right symmetry breaking scale $M_R$ should be at least three times greater than the SUSY breaking 
scale $M_{susy} \sim \mu$. It is interesting to note that this lower bound on $M_R$ is very close to the current 
limit on Left-Right scale $M_{W_R} \geq 2.5 \; \text{TeV}$ \cite{Maiezza:2010ic} leaving the exciting possibility of 
some collider signatures in near future.

\section{Results and Conclusion}
\label{sec:con}
We have discussed in details the recently proposed model by Haba and Ohki \cite{Haba:2011pr} 
where Left-Right symmetry is broken spontaneously and at the same time SUSY is broken 
dynamically in a metastable vacua. Such a model has all the nice features of SUSY 
Left-Right models and also provides a mechanism for SUSY breaking without referring 
to hidden sectors. Such ISS type models have an electric theory (UV free) and a 
magnetic theory (IR free) which are related by the so called Seiberg duality. 
We point out that like many other Left-Right models, this model also does not 
break D-parity globally. There exists two equivalent vacua corresponding to 
left and right sectors of the theory giving rise to the same vacuum energy. 
This will make the accompanying phase transition frustrated leading to the 
formation of domain walls separating different vacua. These domain walls are 
inconsistent with standard cosmology and hence have to disappear somehow at 
least before the BBN. We have discussed how incorporating 
higher dimensional terms in the magnetic theory can make these walls disappear 
provided the Left-Right symmetry breaking scale obeys certain bounds. 
The cornerstone of the construction, condition for the stability of the metastable 
vacua eq. (\ref{stabilitybound})  shows that the scale $\Lambda_m$ has to be at least 
a few orders of magnitude larger than $M_R$.

The natural scale of gauge coupling unification is necessarily as high as $10^{16} \; \text{GeV}$
in this class of models \cite{Kopp:2009xt}\cite{Borah:2010kk} and hence the bounds 
obtained here, $10^{10}-10^{11}$GeV disfavor unification. Further, making use of the 
requirement that the metastable vacuum has a lifetime longer than the age of 
the Universe, we arrive at a lower bound on the parity breaking scale 
$M_R > 3M_{susy} \sim 3 \; \text{TeV}$ which lies just above the present 
lower limit on Left-Right symmetry scale $\sim 2.5 \; \text{TeV}$.
These numbers suggest a natural possibility that the scale $M_R$ is within a few 
orders of magnitude of the TeV scale, leaving open the possibility of signatures 
in future colliders. On the other hand, $\Lambda_m$ is permitted to 
be close to but less than the gravitino bound on the reheat temperature $T_{RH}$.

An alternative is to contemplate the highest possible value permitted by this bound,
since it is close to the intermediate scale of models with low string scale, \cite{LEstrings1}\cite{LEstrings2} 
$10^{11}$GeV. Note that one of 
the scenarios for wall disappearance considered is entirely during "matter dominated" 
era which is generic to string inspired models because of presence 
of a large number of moduli fields that could give such an evolution, alternatively
because  oscillating coherent condensates mimic  this kind of evolution of the scale 
factor. The bound in this case is 
$M_R < 1.3 \times 10^{10} \;\text{GeV}$. But we must then contend with the fact that
the scale is just  above the gravitino bound on reheat temperature after inflation, 
$T_{RH} <10^9 \; \text{GeV}$.
This means that if $M_R$ is as high as allowed by the 
proposed bound and therefore consistent with intermediate scale unification,
then $\Lambda_m$ lies above the reheat temperature $T_{RH}$.

If $\Lambda_m>T_{RH}$, the Universe should already be in the magnetic phase
when inflation ends, since the subsequent evolution of the Universe will not
be able to alter the phase of the SQCD. Setting up such initial conditions is 
more natural to low field inflation than to high field or Planck scale inflation.


\begin{thebibliography}{39}
\expandafter\ifx\csname natexlab\endcsname\relax\def\natexlab#1{#1}\fi
\expandafter\ifx\csname bibnamefont\endcsname\relax
  \def\bibnamefont#1{#1}\fi
\expandafter\ifx\csname bibfnamefont\endcsname\relax
  \def\bibfnamefont#1{#1}\fi
\expandafter\ifx\csname citenamefont\endcsname\relax
  \def\citenamefont#1{#1}\fi
\expandafter\ifx\csname url\endcsname\relax
  \def\url#1{\texttt{#1}}\fi
\expandafter\ifx\csname urlprefix\endcsname\relax\def\urlprefix{URL }\fi
\providecommand{\bibinfo}[2]{#2}
\providecommand{\eprint}[2][]{\url{#2}}

\bibitem[{\citenamefont{Pati and Salam}(1974)}]{Pati:1974yy}
\bibinfo{author}{\bibfnamefont{J.~C.} \bibnamefont{Pati}} \bibnamefont{and}
  \bibinfo{author}{\bibfnamefont{A.}~\bibnamefont{Salam}},
  \bibinfo{journal}{Phys. Rev.} \textbf{\bibinfo{volume}{D10}},
  \bibinfo{pages}{275} (\bibinfo{year}{1974}).

\bibitem[{\citenamefont{Mohapatra and Pati}(1975)}]{Mohapatra:1974gc}
\bibinfo{author}{\bibfnamefont{R.~N.} \bibnamefont{Mohapatra}}
  \bibnamefont{and} \bibinfo{author}{\bibfnamefont{J.~C.} \bibnamefont{Pati}},
  \bibinfo{journal}{Phys. Rev.} \textbf{\bibinfo{volume}{D11}},
  \bibinfo{pages}{2558} (\bibinfo{year}{1975}).

\bibitem[{\citenamefont{Senjanovic and Mohapatra}(1975)}]{Senjanovic:1975rk}
\bibinfo{author}{\bibfnamefont{G.}~\bibnamefont{Senjanovic}} \bibnamefont{and}
  \bibinfo{author}{\bibfnamefont{R.~N.} \bibnamefont{Mohapatra}},
  \bibinfo{journal}{Phys. Rev.} \textbf{\bibinfo{volume}{D12}},
  \bibinfo{pages}{1502} (\bibinfo{year}{1975}).

\bibitem[{\citenamefont{Mohapatra and Marshak}(1980)}]{Mohapatra:1980qe}
\bibinfo{author}{\bibfnamefont{R.~N.} \bibnamefont{Mohapatra}}
  \bibnamefont{and} \bibinfo{author}{\bibfnamefont{R.~E.}
  \bibnamefont{Marshak}}, \bibinfo{journal}{Phys. Rev. Lett.}
  \textbf{\bibinfo{volume}{44}}, \bibinfo{pages}{1316} (\bibinfo{year}{1980}).

\bibitem[{\citenamefont{Deshpande et~al.}(1991)\citenamefont{Deshpande, Gunion,
  Kayser, and Olness}}]{Deshpande:1990ip}
\bibinfo{author}{\bibfnamefont{N.~G.} \bibnamefont{Deshpande}},
  \bibinfo{author}{\bibfnamefont{J.~F.} \bibnamefont{Gunion}},
  \bibinfo{author}{\bibfnamefont{B.}~\bibnamefont{Kayser}}, \bibnamefont{and}
  \bibinfo{author}{\bibfnamefont{F.~I.} \bibnamefont{Olness}},
  \bibinfo{journal}{Phys. Rev.} \textbf{\bibinfo{volume}{D44}},
  \bibinfo{pages}{837} (\bibinfo{year}{1991}).

\bibitem[{\citenamefont{Fukuda et~al.}(2001)}]{Fukuda:2001nk}
\bibinfo{author}{\bibfnamefont{S.}~\bibnamefont{Fukuda}} \bibnamefont{et~al.}
  (\bibinfo{collaboration}{Super-Kamiokande}), \bibinfo{journal}{Phys. Rev.
  Lett.} \textbf{\bibinfo{volume}{86}}, \bibinfo{pages}{5656}
  (\bibinfo{year}{2001}), \eprint{hep-ex/0103033}.

\bibitem[{\citenamefont{Ahmad et~al.}(2002{\natexlab{a}})}]{Ahmad:2002jz}
\bibinfo{author}{\bibfnamefont{Q.~R.} \bibnamefont{Ahmad}} \bibnamefont{et~al.}
  (\bibinfo{collaboration}{SNO}), \bibinfo{journal}{Phys. Rev. Lett.}
  \textbf{\bibinfo{volume}{89}}, \bibinfo{pages}{011301}
  (\bibinfo{year}{2002}{\natexlab{a}}), \eprint{nucl-ex/0204008}.

\bibitem[{\citenamefont{Ahmad et~al.}(2002{\natexlab{b}})}]{Ahmad:2002ka}
\bibinfo{author}{\bibfnamefont{Q.~R.} \bibnamefont{Ahmad}} \bibnamefont{et~al.}
  (\bibinfo{collaboration}{SNO}), \bibinfo{journal}{Phys. Rev. Lett.}
  \textbf{\bibinfo{volume}{89}}, \bibinfo{pages}{011302}
  (\bibinfo{year}{2002}{\natexlab{b}}), \eprint{nucl-ex/0204009}.

\bibitem[{\citenamefont{Bahcall and Pena-Garay}(2004)}]{Bahcall:2004mz}
\bibinfo{author}{\bibfnamefont{J.~N.} \bibnamefont{Bahcall}} \bibnamefont{and}
  \bibinfo{author}{\bibfnamefont{C.}~\bibnamefont{Pena-Garay}},
  \bibinfo{journal}{New J. Phys.} \textbf{\bibinfo{volume}{6}},
  \bibinfo{pages}{63} (\bibinfo{year}{2004}), \eprint{hep-ph/0404061}.

\bibitem[{\citenamefont{Chang et~al.}(1984{\natexlab{a}})\citenamefont{Chang,
  Mohapatra, and Parida}}]{Chang:1983fu}
\bibinfo{author}{\bibfnamefont{D.}~\bibnamefont{Chang}},
  \bibinfo{author}{\bibfnamefont{R.~N.} \bibnamefont{Mohapatra}},
  \bibnamefont{and} \bibinfo{author}{\bibfnamefont{M.~K.}
  \bibnamefont{Parida}}, \bibinfo{journal}{Phys. Rev. Lett.}
  \textbf{\bibinfo{volume}{52}}, \bibinfo{pages}{1072}
  (\bibinfo{year}{1984}{\natexlab{a}}).

\bibitem[{\citenamefont{Chang et~al.}(1984{\natexlab{b}})\citenamefont{Chang,
  Mohapatra, and Parida}}]{Chang:1984uy}
\bibinfo{author}{\bibfnamefont{D.}~\bibnamefont{Chang}},
  \bibinfo{author}{\bibfnamefont{R.~N.} \bibnamefont{Mohapatra}},
  \bibnamefont{and} \bibinfo{author}{\bibfnamefont{M.~K.}
  \bibnamefont{Parida}}, \bibinfo{journal}{Phys. Rev.}
  \textbf{\bibinfo{volume}{D30}}, \bibinfo{pages}{1052}
  (\bibinfo{year}{1984}{\natexlab{b}}).

\bibitem[{\citenamefont{Aulakh et~al.}(1998)\citenamefont{Aulakh, Melfo, and
  Senjanovic}}]{Aulakh:1998nn}
\bibinfo{author}{\bibfnamefont{C.~S.} \bibnamefont{Aulakh}},
  \bibinfo{author}{\bibfnamefont{A.}~\bibnamefont{Melfo}}, \bibnamefont{and}
  \bibinfo{author}{\bibfnamefont{G.}~\bibnamefont{Senjanovic}},
  \bibinfo{journal}{Phys. Rev.} \textbf{\bibinfo{volume}{D57}},
  \bibinfo{pages}{4174} (\bibinfo{year}{1998}), \eprint{hep-ph/9707256}.

\bibitem[{\citenamefont{Aulakh et~al.}(1997)\citenamefont{Aulakh, Benakli, and
  Senjanovic}}]{Aulakh:1997ba}
\bibinfo{author}{\bibfnamefont{C.~S.} \bibnamefont{Aulakh}},
  \bibinfo{author}{\bibfnamefont{K.}~\bibnamefont{Benakli}}, \bibnamefont{and}
  \bibinfo{author}{\bibfnamefont{G.}~\bibnamefont{Senjanovic}},
  \bibinfo{journal}{Phys. Rev. Lett.} \textbf{\bibinfo{volume}{79}},
  \bibinfo{pages}{2188} (\bibinfo{year}{1997}), \eprint{hep-ph/9703434}.

\bibitem[{\citenamefont{Babu and Mohapatra}(2008)}]{Babu:2008ep}
\bibinfo{author}{\bibfnamefont{K.~S.} \bibnamefont{Babu}} \bibnamefont{and}
  \bibinfo{author}{\bibfnamefont{R.~N.} \bibnamefont{Mohapatra}},
  \bibinfo{journal}{Phys. Lett.} \textbf{\bibinfo{volume}{B668}},
  \bibinfo{pages}{404} (\bibinfo{year}{2008}), \eprint{0807.0481}.

\bibitem[{\citenamefont{Patra et~al.}(2009)\citenamefont{Patra, Sarkar, Sarkar,
  and Yajnik}}]{Patra:2009wc}
\bibinfo{author}{\bibfnamefont{S.}~\bibnamefont{Patra}},
  \bibinfo{author}{\bibfnamefont{A.}~\bibnamefont{Sarkar}},
  \bibinfo{author}{\bibfnamefont{U.}~\bibnamefont{Sarkar}}, \bibnamefont{and}
  \bibinfo{author}{\bibfnamefont{U.}~\bibnamefont{Yajnik}},
  \bibinfo{journal}{Phys. Lett.} \textbf{\bibinfo{volume}{B679}},
  \bibinfo{pages}{386} (\bibinfo{year}{2009}), \eprint{0905.3220}.

\bibitem[{\citenamefont{Kibble}(1980)}]{Kibble:1980mv}
\bibinfo{author}{\bibfnamefont{T.~W.~B.} \bibnamefont{Kibble}},
  \bibinfo{journal}{Phys. Rept.} \textbf{\bibinfo{volume}{67}},
  \bibinfo{pages}{183} (\bibinfo{year}{1980}).

\bibitem[{\citenamefont{Hindmarsh and Kibble}(1995)}]{Hindmarsh:1994re}
\bibinfo{author}{\bibfnamefont{M.~B.} \bibnamefont{Hindmarsh}}
  \bibnamefont{and} \bibinfo{author}{\bibfnamefont{T.~W.~B.}
  \bibnamefont{Kibble}}, \bibinfo{journal}{Rept. Prog. Phys.}
  \textbf{\bibinfo{volume}{58}}, \bibinfo{pages}{477} (\bibinfo{year}{1995}),
  \eprint{hep-ph/9411342}.

\bibitem[{\citenamefont{Rai and Senjanovic}(1994)}]{Rai:1992xw}
\bibinfo{author}{\bibfnamefont{B.}~\bibnamefont{Rai}} \bibnamefont{and}
  \bibinfo{author}{\bibfnamefont{G.}~\bibnamefont{Senjanovic}},
  \bibinfo{journal}{Phys. Rev.} \textbf{\bibinfo{volume}{D49}},
  \bibinfo{pages}{2729} (\bibinfo{year}{1994}), \eprint{hep-ph/9301240}.

\bibitem[{\citenamefont{Lew and Riotto}(1993)}]{Lew:1993yt}
\bibinfo{author}{\bibfnamefont{H.}~\bibnamefont{Lew}} \bibnamefont{and}
  \bibinfo{author}{\bibfnamefont{A.}~\bibnamefont{Riotto}},
  \bibinfo{journal}{Phys. Lett.} \textbf{\bibinfo{volume}{B309}},
  \bibinfo{pages}{258} (\bibinfo{year}{1993}), \eprint{hep-ph/9304203}.
  
  \bibitem[{\citenamefont{Abel et~al.}(1995)\citenamefont{Abel, Sarkar, and
  White}}]{Abel:1995wk}
\bibinfo{author}{\bibfnamefont{S.~A.} \bibnamefont{Abel}},
  \bibinfo{author}{\bibfnamefont{S.}~\bibnamefont{Sarkar}}, \bibnamefont{and}
  \bibinfo{author}{\bibfnamefont{P.~L.} \bibnamefont{White}},
  \bibinfo{journal}{Nucl. Phys.} \textbf{\bibinfo{volume}{B454}},
  \bibinfo{pages}{663} (\bibinfo{year}{1995}), \eprint{hep-ph/9506359}.
 
\bibitem[{\citenamefont{Mishra and Yajnik}(2010)}]{Mishra:2009mk}
\bibinfo{author}{\bibfnamefont{S.}~\bibnamefont{Mishra}} \bibnamefont{and}
  \bibinfo{author}{\bibfnamefont{U.~A.} \bibnamefont{Yajnik}},
  \bibinfo{journal}{Phys. Rev.} \textbf{\bibinfo{volume}{D81}},
  \bibinfo{pages}{045010} (\bibinfo{year}{2010}), \eprint{0911.1578}.

\bibitem[{\citenamefont{Borah and Mishra}(2011)}]{Borah:2011qq}
\bibinfo{author}{\bibfnamefont{D.}~\bibnamefont{Borah}} \bibnamefont{and}
  \bibinfo{author}{\bibfnamefont{S.}~\bibnamefont{Mishra}}
  (\bibinfo{year}{2011}), \eprint{1105.5006}.

\bibitem{Mishra:2008be}
  S.~Mishra, U.~A.~Yajnik and A.~Sarkar,
  Phys.\ Rev.\  D {\bf 79}, 065038 (2009)

\bibitem[{\citenamefont{Intriligator et~al.}(2006)\citenamefont{Intriligator,
  Seiberg, and Shih}}]{Intriligator:2006dd}
\bibinfo{author}{\bibfnamefont{K.~A.} \bibnamefont{Intriligator}},
  \bibinfo{author}{\bibfnamefont{N.}~\bibnamefont{Seiberg}}, \bibnamefont{and}
  \bibinfo{author}{\bibfnamefont{D.}~\bibnamefont{Shih}},
  \bibinfo{journal}{JHEP} \textbf{\bibinfo{volume}{04}}, \bibinfo{pages}{021}
  (\bibinfo{year}{2006}), \eprint{hep-th/0602239}.

\bibitem[{\citenamefont{Abel and Khoze}(2007)}]{Abel:2007uq}
\bibinfo{author}{\bibfnamefont{S.~A.} \bibnamefont{Abel}} \bibnamefont{and}
  \bibinfo{author}{\bibfnamefont{V.~V.} \bibnamefont{Khoze}}
  (\bibinfo{year}{2007}), \eprint{hep-ph/0701069}.

\bibitem{Koschade:2009qu}
  D.~Koschade, M.~McGarrie, S.~Thomas,
  JHEP {\bf 1002}, 100 (2010).
  [arXiv:0909.0233 [hep-ph]].

\bibitem[{\citenamefont{Haba and Ohki}(2011)}]{Haba:2011pr}
\bibinfo{author}{\bibfnamefont{N.}~\bibnamefont{Haba}} \bibnamefont{and}
  \bibinfo{author}{\bibfnamefont{H.}~\bibnamefont{Ohki}}
  (\bibinfo{year}{2011}), \eprint{1104.5405}.

\bibitem{Borah:2010kk}
  D.~Borah, U.~A.~Yajnik,
  ``Supersymmetric Left-Right models with Gauge Coupling Unification and Fermion Mass Universality,''
  Phys.\ Rev.\  {\bf D83}, 095004 (2011).
  [arXiv:1010.6289 [hep-ph]].

\bibitem{LEstrings1}
C.~P.~Burgess, L.~E.~Ib\'a\~nez and F.~Quevedo,
 {\it Phys.~Lett.} {\bf B447}, 257 (1999) [hep-ph/{9810535}] 
\bibitem{LEstrings2}
I.~Antoniadis and K.~Benakli,
  Int.\ J.\ Mod.\ Phys.\  A {\bf 15} (2000) 4237
  [arXiv:hep-ph/0007226].

\bibitem[{\citenamefont{Sarkar et~al.}(2008)\citenamefont{Sarkar, Abhishek, and
  Yajnik}}]{Sarkar:2007er}
\bibinfo{author}{\bibfnamefont{A.}~\bibnamefont{Sarkar}},
  \bibinfo{author}{\bibnamefont{Abhishek}}, \bibnamefont{and}
  \bibinfo{author}{\bibfnamefont{U.~A.} \bibnamefont{Yajnik}},
  \bibinfo{journal}{Nucl. Phys.} \textbf{\bibinfo{volume}{B800}},
  \bibinfo{pages}{253} (\bibinfo{year}{2008}), \eprint{0710.5410}.

\bibitem[{\citenamefont{Dev and Mohapatra}(2010)}]{Dev:2009aw}
\bibinfo{author}{\bibfnamefont{P.~S.~B.} \bibnamefont{Dev}} \bibnamefont{and}
  \bibinfo{author}{\bibfnamefont{R.~N.} \bibnamefont{Mohapatra}},
  \bibinfo{journal}{Phys. Rev.} \textbf{\bibinfo{volume}{D81}},
  \bibinfo{pages}{013001} (\bibinfo{year}{2010}), \eprint{0910.3924}.

\bibitem[{\citenamefont{Borah et~al.}(2011)\citenamefont{Borah, Patra, and
  Sarkar}}]{Borah:2010zq}
\bibinfo{author}{\bibfnamefont{D.}~\bibnamefont{Borah}},
  \bibinfo{author}{\bibfnamefont{S.}~\bibnamefont{Patra}}, \bibnamefont{and}
  \bibinfo{author}{\bibfnamefont{U.}~\bibnamefont{Sarkar}},
  \bibinfo{journal}{Phys. Rev.} \textbf{\bibinfo{volume}{D83}},
  \bibinfo{pages}{035007} (\bibinfo{year}{2011}), \eprint{1006.2245}.

\bibitem[{\citenamefont{Borah}(2011)}]{Borah:2011zz}
\bibinfo{author}{\bibfnamefont{D.}~\bibnamefont{Borah}}, \bibinfo{journal}{Int.
  J. Mod. Phys.} \textbf{\bibinfo{volume}{A26}}, \bibinfo{pages}{1305}
  (\bibinfo{year}{2011}).

\bibitem[{\citenamefont{Vilenkin}(1985)}]{Vilenkin:1984ib}
\bibinfo{author}{\bibfnamefont{A.}~\bibnamefont{Vilenkin}},
  \bibinfo{journal}{Phys. Rept.} \textbf{\bibinfo{volume}{121}},
  \bibinfo{pages}{263} (\bibinfo{year}{1985}).

\bibitem[{\citenamefont{Kawasaki and Takahashi}(2005)}]{Kawasaki:2004rx}
\bibinfo{author}{\bibfnamefont{M.}~\bibnamefont{Kawasaki}} \bibnamefont{and}
  \bibinfo{author}{\bibfnamefont{F.}~\bibnamefont{Takahashi}},
  \bibinfo{journal}{Phys. Lett.} \textbf{\bibinfo{volume}{B618}},
  \bibinfo{pages}{1} (\bibinfo{year}{2005}), \eprint{hep-ph/0410158}.

\bibitem[{\citenamefont{Seiberg}(1995)}]{Seiberg:1994pq}
\bibinfo{author}{\bibfnamefont{N.}~\bibnamefont{Seiberg}},
  \bibinfo{journal}{Nucl. Phys.} \textbf{\bibinfo{volume}{B435}},
  \bibinfo{pages}{129} (\bibinfo{year}{1995}), \eprint{hep-th/9411149}.

\bibitem[{\citenamefont{Fukugita and Yanagida}(1986)}]{Fukugita:1986hr}
\bibinfo{author}{\bibfnamefont{M.}~\bibnamefont{Fukugita}} \bibnamefont{and}
  \bibinfo{author}{\bibfnamefont{T.}~\bibnamefont{Yanagida}},
  \bibinfo{journal}{Phys. Lett.} \textbf{\bibinfo{volume}{B174}},
  \bibinfo{pages}{45} (\bibinfo{year}{1986}).

\bibitem[{\citenamefont{Guo et~al.}(2011)\citenamefont{Guo, Schwarz, and
  Zhang}}]{Guo:2010mm}
\bibinfo{author}{\bibfnamefont{Z.-K.} \bibnamefont{Guo}},
  \bibinfo{author}{\bibfnamefont{D.~J.} \bibnamefont{Schwarz}},
  \bibnamefont{and} \bibinfo{author}{\bibfnamefont{Y.-Z.} \bibnamefont{Zhang}},
  \bibinfo{journal}{Phys. Rev.} \textbf{\bibinfo{volume}{D83}},
  \bibinfo{pages}{083522} (\bibinfo{year}{2011}), \eprint{1008.5258}.

\bibitem[{\citenamefont{Boyle et~al.}(2006)\citenamefont{Boyle, Steinhardt, and
  Turok}}]{Boyle:2005ug}
\bibinfo{author}{\bibfnamefont{L.~A.} \bibnamefont{Boyle}},
  \bibinfo{author}{\bibfnamefont{P.~J.} \bibnamefont{Steinhardt}},
  \bibnamefont{and} \bibinfo{author}{\bibfnamefont{N.}~\bibnamefont{Turok}},
  \bibinfo{journal}{Phys. Rev. Lett.} \textbf{\bibinfo{volume}{96}},
  \bibinfo{pages}{111301} (\bibinfo{year}{2006}), \eprint{astro-ph/0507455}.

\bibitem[{\citenamefont{Komatsu et~al.}(2009)}]{Komatsu:2008hk}
\bibinfo{author}{\bibfnamefont{E.}~\bibnamefont{Komatsu}} \bibnamefont{et~al.}
  (\bibinfo{collaboration}{WMAP}), \bibinfo{journal}{Astrophys. J. Suppl.}
  \textbf{\bibinfo{volume}{180}}, \bibinfo{pages}{330} (\bibinfo{year}{2009}),
  \eprint{0803.0547}.

\bibitem[{\citenamefont{Coleman}(1977)}]{Coleman:1977py}
\bibinfo{author}{\bibfnamefont{S.~R.} \bibnamefont{Coleman}},
  \bibinfo{journal}{Phys. Rev.} \textbf{\bibinfo{volume}{D15}},
  \bibinfo{pages}{2929} (\bibinfo{year}{1977}).

\bibitem[{\citenamefont{Duncan and Jensen}(1992)}]{Duncan:1992ai}
\bibinfo{author}{\bibfnamefont{M.~J.} \bibnamefont{Duncan}} \bibnamefont{and}
  \bibinfo{author}{\bibfnamefont{L.~G.} \bibnamefont{Jensen}},
  \bibinfo{journal}{Phys. Lett.} \textbf{\bibinfo{volume}{B291}},
  \bibinfo{pages}{109} (\bibinfo{year}{1992}).

\bibitem[{\citenamefont{Maiezza et~al.}(2010)\citenamefont{Maiezza, Nemevsek,
  Nesti, and Senjanovic}}]{Maiezza:2010ic}
\bibinfo{author}{\bibfnamefont{A.}~\bibnamefont{Maiezza}},
  \bibinfo{author}{\bibfnamefont{M.}~\bibnamefont{Nemevsek}},
  \bibinfo{author}{\bibfnamefont{F.}~\bibnamefont{Nesti}}, \bibnamefont{and}
  \bibinfo{author}{\bibfnamefont{G.}~\bibnamefont{Senjanovic}},
  \bibinfo{journal}{Phys. Rev.} \textbf{\bibinfo{volume}{D82}},
  \bibinfo{pages}{055022} (\bibinfo{year}{2010}), \eprint{1005.5160}.

\bibitem[{\citenamefont{Kopp et~al.}(2010)\citenamefont{Kopp, Lindner, Niro,
  and Underwood}}]{Kopp:2009xt}
\bibinfo{author}{\bibfnamefont{J.}~\bibnamefont{Kopp}},
  \bibinfo{author}{\bibfnamefont{M.}~\bibnamefont{Lindner}},
  \bibinfo{author}{\bibfnamefont{V.}~\bibnamefont{Niro}}, \bibnamefont{and}
  \bibinfo{author}{\bibfnamefont{T.~E.~J.} \bibnamefont{Underwood}},
  \bibinfo{journal}{Phys. Rev.} \textbf{\bibinfo{volume}{D81}},
  \bibinfo{pages}{025008} (\bibinfo{year}{2010}), \eprint{0909.2653}.

\end{thebibliography}

\end{document}